\documentclass[floatfix,%
 reprint,fleqn,%
superscriptaddress,%
 amsmath,amssymb,
 aps,
 prl
]{revtex4-2}

\usepackage{amssymb,amsmath,amstext}
\usepackage{graphicx}
\graphicspath{{pictures/}}
\usepackage{bm}
\usepackage{xcolor}
\usepackage{physics}
\usepackage{dcolumn}
\usepackage[caption=false]{subfig}
\usepackage{hyperref}
\usepackage{orcidlink} 

\newcommand{\Ntb}{N_{\mathrm{TB}}}
\newcommand{\Nsb}{N_{\mathrm{SB}}}
\newcommand{\Nsbt}{N_{\mathrm{SBT}}}
\newcommand{\Ssbt}{\mathrm{S}_{\mathrm{SBT}}}
\newcommand{\Smsb}{\mathrm{S}_{\mathrm{SB}}}

\newif\ifincludeSM
\includeSMtrue

\begin{document}

\preprint{APS/123-QED}

\title{Insight into ordering at nematic twist-bend interfaces}%

\author{Szymon Drzazga\orcidlink{0009-0009-6068-3097}}%
\affiliation{$^{1}$Institute of Theoretical Physics and  Mark Kac Center 
for Complex Systems Research, Jagiellonian University, \L{}ojasiewicza 11, 30-348 Krak\'ow, Poland}%
\email{szymon.drzazga@student.uj.edu.pl}%
\author{Piotr Kubala\orcidlink{0000-0001-7008-4498}}%
\affiliation{$^{1}$Institute of Theoretical Physics and  Mark Kac Center 
for Complex Systems Research, Jagiellonian University, \L{}ojasiewicza 11, 30-348 Krak\'ow, Poland}%
\email{piotr.kubala@doctoral.uj.edu.pl}%

\author{Lech Longa\orcidlink{0000-0002-4918-6518}}%
\affiliation{$^{1}$Institute of Theoretical Physics and  Mark Kac Center 
for Complex Systems Research, Jagiellonian University, \L{}ojasiewicza 11, 30-348 Krak\'ow, Poland}%
\email{lech.longa@uj.edu.pl}%

\date{\today}

\begin{abstract}
Twist--bend nematics formed by achiral particles support heliconical
domains of opposite handedness. Using Monte Carlo and molecular
dynamics simulations of repulsive bent particles, we study the
interface between two such domains. Rather than a gradually
untwisting texture, we find a density-modulated splay--bend--twist
structure. The density modulation is along the helix axis with a
period of approximately half the bulk pitch, while the twist is
locally enhanced in magnitude, alternates in sign, and vanishes
only on an undulating surface. An explicit director interpolation shows how gradients along the
helix axis and across the interface combine to produce an
undulating zero-twist surface.

\end{abstract}

\maketitle

\textit{Introduction---}Recent discoveries have revealed new forms of
polar and spatially modulated nematic order. These include the
ferroelectric nematic, $N_{\mathrm{F}}$
\cite{Nishikawa2017,PhysRevLett.124.037801,Chen2020},
its heliconical counterpart, $N_{\mathrm{TBF}}$
\cite{Karcz2024,Nishikawa2024},
the twist--bend nematic, $\Ntb$
\cite{Cestari2011,Borshch2013,Chen2013},
and the splay--bend nematic, $\Nsb$
\cite{Fernandez-Rico2020,Kotni2022}.

\begin{figure}
    \centering
    \includegraphics[width=\linewidth]{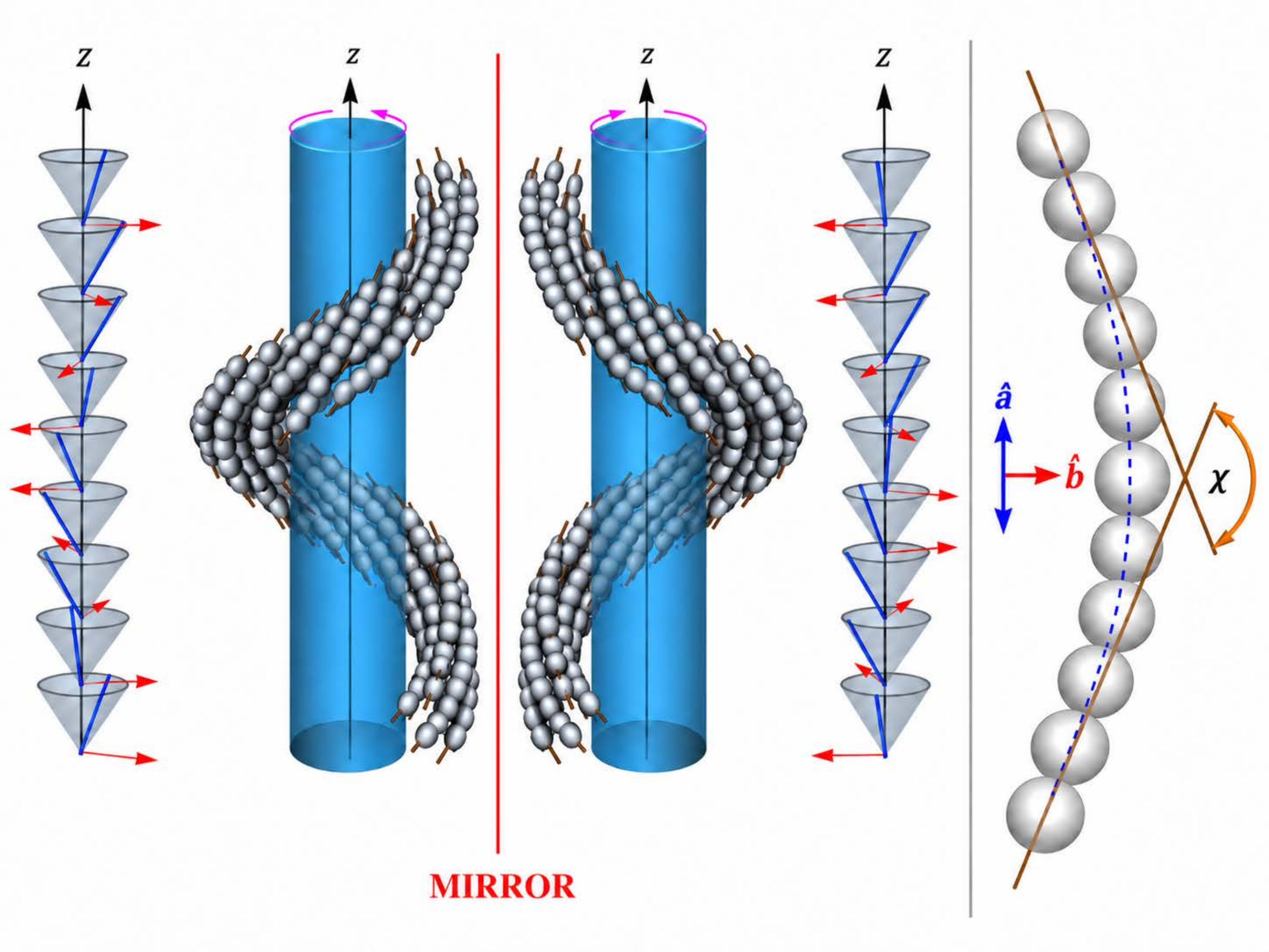}
    \caption{The central pair illustrates mirror-related heliconical
$\Ntb$ states of opposite handedness, with their helix axes along $z$:
blue lines show the local director $\hat{\bm n}$, red arrows the local
polar field $\bm p$, and the gray cones the constant director tilt
relative to the helix axis. The rightmost panel shows the $C_{2v}$-symmetric model molecule,
constructed from eleven identical tangent spheres whose centers
are equally spaced along a circular arc. The bend angle $\chi$,
molecular long axis $\hat{\bm a}$, and polar twofold axis
$\hat{\bm b}$ are indicated. Larger $\chi$ corresponds to smaller molecular
curvature.}
    \label{fig:molecule_model}
\end{figure}

In the $\Ntb$ phase, the apolar unit director field $\hat{\bm n}(\bm r)$
forms a heliconical texture of constant tilt about the helix axis
(Fig.~\ref{fig:molecule_model}), with a pitch typically on the order of 
$10\,$nm and no long-range positional order of the molecular centers of mass.
 For achiral particles in the absence of a
chiral bias, the bulk free energy is invariant under reflection,
which interchanges the left- and right-handed heliconical
states. These non mirror-symmetric states consequently have equal bulk free-energy
density.

Local polar order is described by the dimensionless field
$\bm p(\bm r)$, defined as the local average of the molecular
polar twofold axis $\hat{\bm b}$
(Fig.~\ref{fig:molecule_model}). This field couples to director
bend and can be taken as a primary order parameter of the
$\Ntb$ phase~\cite{Shamid2013statistical,Longa2020}.
In the ideal bulk heliconical state, its direction defines
the local $C_2$ symmetry axis and is perpendicular to both
the director and the helix axis.

For a helix of wave vector $q_0\hat{\bm z}$, $q_0>0$, and pitch
$\lambda=2\pi/q_0$, with $\alpha=q_0z+\phi_0$ for an arbitrary phase
$\phi_0$, cone angle $0<\theta_0<\pi/2$ and polar amplitude $p_0>0$,
the ideal bulk fields of the two mirror-related monodomains are
\[
\begin{aligned}
\hat{\bm n}_{\pm}(z)
&=(\sin\theta_0\cos\alpha,
   \ \pm\sin\theta_0\sin\alpha,\ \cos\theta_0),\\
\bm p_{\pm}(z)
&=p_0(\sin\alpha,\ \mp\cos\alpha,\ 0).
\end{aligned}
\]
Local twist, measured by the Oseen--Frank invariant
$\tau(\bm r)\equiv\hat{\bm n}\cdot(\nabla\times\hat{\bm n})$, has
opposite signs in the two domains,
$\tau_\pm=\mp q_0\sin^2\theta_0$, while $\bm p_\pm$ averages to zero
over one pitch. In the simplest planar $\Nsb$ texture, by contrast, the
director oscillates in a single plane about a mean axis, generating
alternating regions of splay and bend~\cite{Dozov2001}. Clearly, the twist of
such a planar one-dimensional field vanishes identically.

Excluded-volume effects alone can stabilize $\Ntb$, as shown by a
generalized Onsager theory and molecular dynamics (MD) for hard bent
particles~\cite{GrecoFerrariniPRL}. Subsequent Monte Carlo (MC) and MD simulations determined
the full phase diagram of this model as a function of
bend angle and packing fraction~\cite{Kubala2022}.

A natural question is how heliconical $\Ntb$ textures of
opposite handedness self-organize across a domain boundary.
Optical measurements on CB7CB suggest splay--bend nematic
order in the cores of such walls~\cite{Meyer2015}, but leave
their microscopic structure unresolved. Here we determine the interfacial structure in a model of
repulsive bent particles~\cite{Drzazga2026PRE} using Monte Carlo
(MC) and molecular dynamics (MD) simulations. We show that
the interface differs markedly from the anticipated nematic splay--bend
texture, instead exhibiting a density-modulated
splay--bend--twist structure.

\textit{Model---}
Each molecule is a rigid assembly of eleven identical spheres of
diameter $\sigma$, with neighboring spheres tangent and their centers
equally spaced along a circular arc. The unit vector $\hat{\bm a}$ is
parallel to the chord joining the terminal centers, and $\hat{\bm b}$ is
the twofold axis directed toward the convex side
(Fig.~\ref{fig:molecule_model}). The bend angle $\chi$, measured between
the tangents to the arc at the terminal centers, both directed inward
along the chain, runs from $\chi=180^\circ$ (straight chain) to
$\chi=0^\circ$ (semicircle); for $0\leq\chi<\pi$ the angular separation
of neighboring centers is $\delta=(\pi-\chi)/10$ and the arc radius is
$R=\sigma/[2\sin(\delta/2)]$. For all bend angles we use the fixed
reference length $L_{\mathrm{mol}}=11\sigma$, the axial extent of the
straight chain, and the packing fraction
$\eta=Nv_{\mathrm{mol}}/V$ with $v_{\mathrm{mol}}=11\pi\sigma^3/6$,
where $N$ is the number of molecules and $V$ is the box volume.

The MC model uses hard-sphere exclusion, Being athermal, its equation of
state depends only on the dimensionless pressure
$\beta P\sigma^3=P\sigma^3/(k_BT)$~\cite{Frenkel1987,Mederos2014}.
In MD, all pairs of spheres belonging to different molecules interact
through the Weeks--Chandler--Andersen
potential~\cite{WCA1971,WCA1983}, a Lennard-Jones interaction of length
scale $\sigma$ and energy scale $\epsilon$ truncated and shifted at its
minimum, $r_c=2^{1/6}\sigma$. This introduces the reduced temperature $T^*=k_BT/\epsilon$,
on which the structure depends only weakly~\cite{Heyes2006}.

A WCA sphere with length parameter $\sigma$ can be assigned
a temperature-dependent effective hard-sphere diameter
$\sigma_{\mathrm{eff}}$, given by
\begin{equation}\label{eq:sigma_eff}
\frac{\sigma_{\mathrm{eff}}}{\sigma}
=
\frac{0.3837+1.068\,T^*}{0.4293+T^*}.
\end{equation}
At the MD temperature $T^*=1$, this gives
$\sigma_{\mathrm{eff}}\approx1.0157\,\sigma$,
slightly larger than the WCA length parameter~\cite{Heyes2006}. 
We therefore multiply the molecular volume
$v_{\mathrm{mol}}$ used to calculate the MD packing fraction
by $(\sigma_{\mathrm{eff}}/\sigma)^3\approx1.047$.
The distances between sphere centers remain unchanged.
The small overlaps between adjacent effective spheres are
neglected, following Ref.~\cite{Drzazga2026PRE}.
This rescaling applies only to the volume used in the packing
fraction. All other lengths, including $L_{\mathrm{mol}}$,
retain their nominal values defined using $\sigma$.

\begin{figure}[htb]
    \centering
    \includegraphics[width=0.7\linewidth]{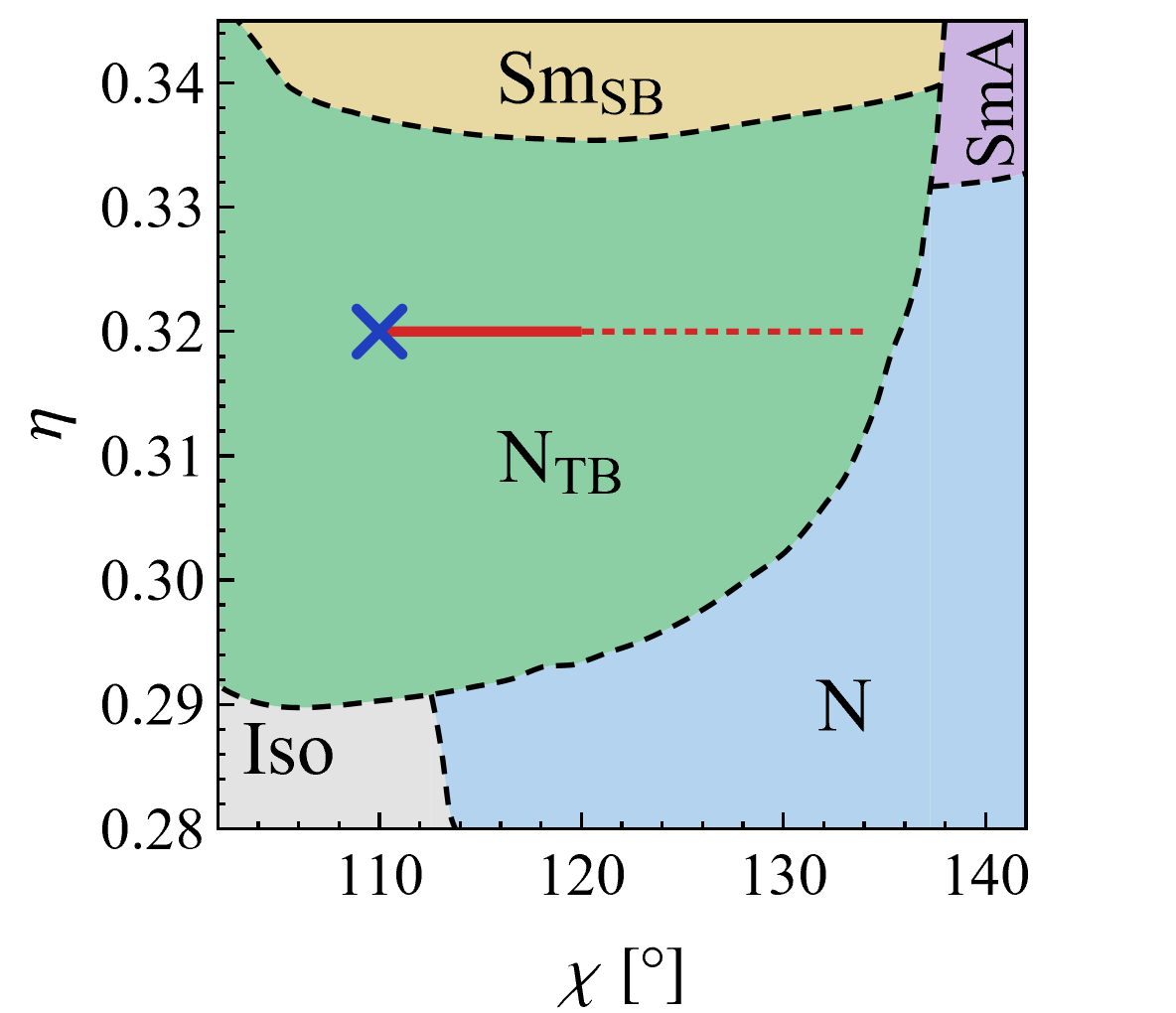}
    \caption{
The phase diagram of the athermal hard-sphere model,
adapted from the $NPT$ MC results of Ref.~\cite{Kubala2022},
is shown over the relevant range of bend angles $\chi$
and packing fractions $\eta$. The MD packing fractions
are mapped onto the same $\eta$ scale using the effective
WCA diameter given by Eq.~\eqref{eq:sigma_eff}.
The blue cross marks the state point studied in detail,
$\chi=110^\circ$ and $\eta=0.320$. The solid red line
indicates the range over which both domains persisted
for $5\times10^7$ MD time steps. The dashed line indicates
the tested states in which one domain disappeared for a given system size.
At $\chi=110^\circ$, the $\Ntb$ phase is bounded on the
high-density side by the splay--bend smectic phase $\Smsb$.
}
    \label{fig:pd}
\end{figure}

We use the corrected packing fractions to plot the MD state
points on the $\eta$ axis of the hard-sphere MC phase diagram
(Fig.~\ref{fig:pd}). In Ref.~\cite{Kubala2022}, the two models
were compared directly at equal nominal packing fraction
$\eta$ and yielded phase
boundaries that were indistinguishable within the estimated
uncertainties.
The effective-diameter correction is included in the present
comparison but was not applied in the earlier study.
The MD state points are characterized by $T^*=1$ and the
mean packing fraction $\eta$, obtained by time averaging
the instantaneous packing fraction in the $NPT$ ensemble.
For the athermal MC simulations, we choose the reduced
pressure $\beta P\sigma^3$ to match this mean packing fraction.

The interfacial maps presented below correspond to
$\chi=110^\circ$ and $\eta=0.320$, well within the $\Ntb$
stability region (Fig.~\ref{fig:pd}). We also investigate
how the interfacial structure depends on the bend angle $\chi$.

\textit{Simulations---} We studied interfaces between domains of opposite handedness
in the stable $\Ntb$ phase. Both MC and MD simulations were
performed in the $NPT$ ensemble with periodic boundary conditions
in all three directions. The MC simulations used
RAMPACK~\cite{RAMPACK} with 6000 molecules per domain,
$12\,000$ in total. Each run comprised $5\times10^8$ MC cycles.
The MD simulations used LAMMPS~\cite{LAMMPS}
(version 14 December 2021) with a reduced time step of 0.005.
The total linear momentum was reset every $10^6$ steps.

The bulk pitch was estimated from a single-domain MD system
of 9000 molecules in a box accommodating two helical periods
along $\hat{\bm z}$. Following a relaxation run of
$6\times10^7$ steps, we obtained
$\lambda\simeq3.2\,L_{\mathrm{mol}}$.
The two-domain MD system contained 9000 molecules per domain,
giving $18\,000$ molecules in total.

We constructed the second domain by reflecting the first
in a plane normal to $\hat{\bm x}$, thereby reversing its
handedness, and then translating it by $\lambda/2$ along
$\hat{\bm z}$. In simulations initialized without this translation, the two
domains spontaneously developed a relative helical phase
shift of $\pi$ after a finite number of MC cycles or MD time
steps. This shift can lower the interfacial free energy by reducing
transverse splay in favor of twist. This mechanism is favored
when the splay elastic constant exceeds the twist elastic
constant, a condition commonly encountered in nematics.
We therefore imposed this
half-pitch shift when preparing the initial two-domain
system from the equilibrated bulk configuration.
For the ideal bulk director fields, the combined reflection
and translation preserve $n_x$ and $n_z$ and reverse $n_y$.
The two-domain MD system was then relaxed for $10^7$ time
steps, followed by a production run of $5\times10^7$ time
steps during which 2500 snapshots were collected. 
At equilibrium, the mean interfacial structure was
translationally invariant along $y$, with spatial variations
confined to the $xz$ plane. We therefore present the director
and polarization fields in $xz$ cross sections.

For statistical analysis, molecular centers were binned on a $150\times150$ grid in the $xz$
plane, and fields were averaged over $y$ and over the sampling interval.
With $\mathsf Q=\langle\tfrac32\hat{\bm a}\otimes\hat{\bm a}
-\tfrac12\mathsf I\rangle$ and $\bm p=\langle\hat{\bm b}\rangle$, the
director is the principal eigenvector of $\mathsf Q$, oriented with
$n_z>0$. The density is normalized by its bulk value, and spatial
derivatives were evaluated using the standard central difference method.

\begin{figure}
\centering
\includegraphics[width=\linewidth]{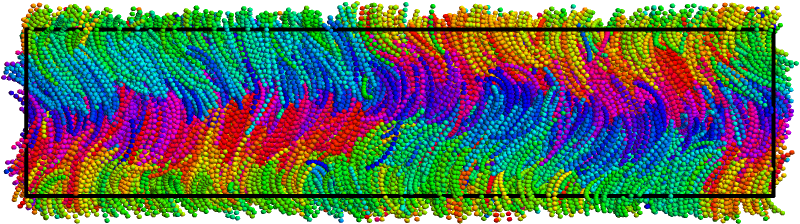}
\caption{MC snapshot of two adjacent $\Ntb$ domains of opposite
handedness. Hue represents the azimuth of the molecular polar axis
projected onto the plane perpendicular to the helix axis.}
\label{fig:domains_snap}
\end{figure}

\begin{figure}
\centering
\includegraphics[width=\linewidth]{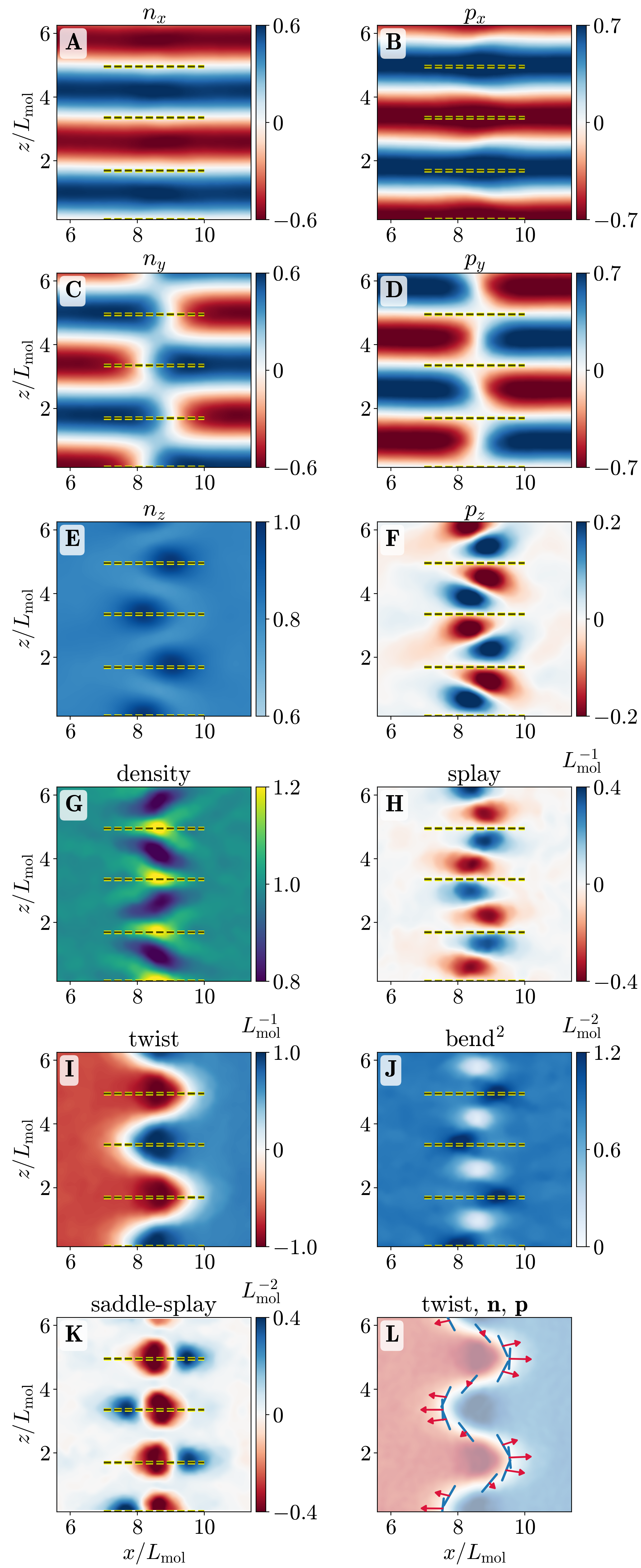}
\caption{MD fields near one of the two chirality-reversing interfaces
at $\chi=110^\circ$ and $\eta=0.320$, averaged over $y$ and
2500 snapshots from $5\times10^7$ production steps. The mean interface
normal is $\hat{\bm x}$ and the helix axis is $\hat{\bm z}$.
Panels A--F show the components of $\hat{\bm n}$ and $\bm p$.
Panels G--K show the normalized density, splay, twist, squared bend,
and saddle-splay. Panel L shows the twist map with director segments
and polar-order arrows; the white contour marks $\tau=0$.
Markers indicate the positions of the density maxima along the helix
axis and the width of the interfacial structure along $\hat{\bm x}$.}
\label{fig:MD_maps}
\end{figure}

\begin{figure}
\centering
\includegraphics[width=\linewidth]{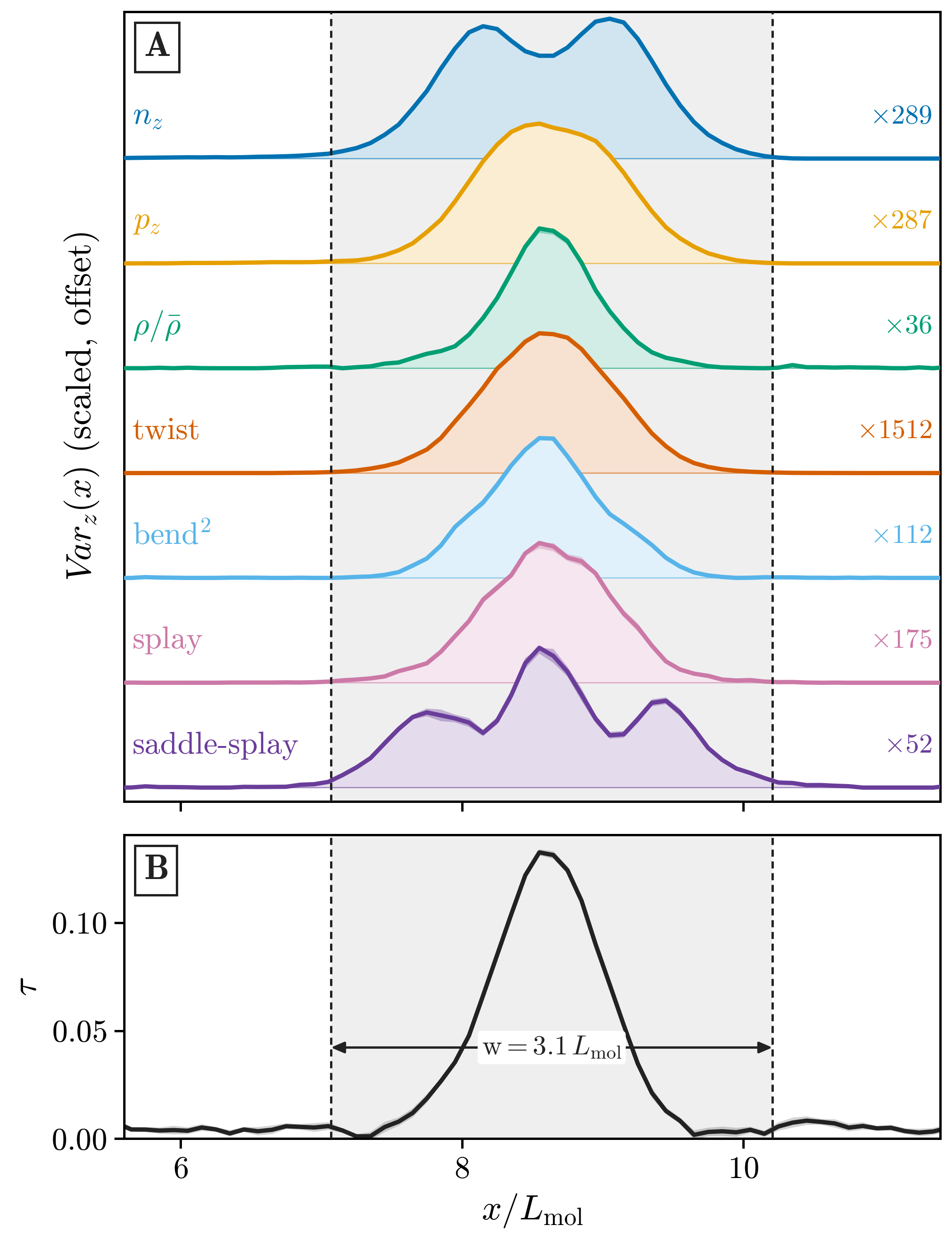}
\caption{
Spatial extent and density modulation of the domain wall.
(A) The variance along $z$ is shown as a function of $x$
for $n_z$, $p_z$, the normalized density $\rho/\bar{\rho}$,
twist, squared bend, splay, and saddle-splay.
These quantities are uniform in the ideal bulk $\Ntb$ state.
Each quantity is averaged over $y$ before its variance along
$z$ is calculated. The profiles are normalized as
$[\mathrm{Var}_z(x)-\mathrm{Var}_b]/
[\mathrm{Var}_{\max}-\mathrm{Var}_b]$
and offset vertically. Here $\mathrm{Var}_b$ and
$\mathrm{Var}_{\max}$ denote the bulk and maximum variances
of the corresponding quantity. The annotations on the right
give the peak-to-bulk ratios
$\mathrm{Var}_{\max}/\mathrm{Var}_b$.
The colored fills highlight the profiles above their
respective bulk baselines.
(B) The local smectic order parameter is
$\tau(x)=
\left|\sum_{j,k}\rho_{jk}(x)e^{2\pi i z_k/d}\right|
/\sum_{j,k}\rho_{jk}(x)$,
where $\rho_{jk}(x)$ is the density in bin $(y_j,z_k)$
of the slab at $x$ and $d=1.6\,L_{\mathrm{mol}}$ is the
layer spacing. Narrow shaded bands around the curves
indicate $\pm1$ jackknife standard error.
The vertical dashed lines delimit the interfacial region,
highlighted in gray. Outside this region, all normalized
variance profiles in (A) are below 0.05.
The resulting wall width is $w=3.1\,L_{\mathrm{mol}}$,
as indicated by the horizontal arrow in (B).
}
\label{fig:MD_variance_plots}
\end{figure}

\textit{Interfacial structure---} As already mentioned, 
the mean local  director and polar fields are independent of $y$,
so Fig.~\ref{fig:MD_maps} presents them in $xz$ cross sections.
Far from the interface, the director approaches the bulk
heliconical states $\hat{\bm n}_\pm$, and the polar field
recovers the corresponding bulk profiles. The twist has
opposite signs in the two domains. Across the wall, the
$n_y$ and $p_y$ modulations reverse sign, whereas those of
$n_x$ and $p_x$ do not. The interfacial region has a width
$w\approx3.1\,L_{\mathrm{mol}}$
(Fig.~\ref{fig:MD_variance_plots}). Within this region,
all fields shown deviate from their corresponding bulk
profiles.

The zero of $n_y$ does not lie on the nominal wall midplane
$x/L_{\mathrm{mol}}\simeq8.5$. It is displaced by an amount that
varies with $z$ over each pitch [Fig.~\ref{fig:MD_maps}(C)], so the
mean director on that plane is only approximately planar. To this
accuracy it lies in the $xz$ plane and undergoes a periodic
orientational modulation with period $\lambda$,
aligning with the helix axis twice per pitch.
By contrast, the polar field lies in the $xz$ plane
and points predominantly along $\pm\hat{\bm x}$. Its average over one
pitch vanishes. The director profile along the midplane thus has,
to the same accuracy, the planar form associated with splay--bend
order, consistent with the interpretation of optical
measurements~\cite{Meyer2015}. Splay is localized at the
wall and vanishes in the ideal bulk $\Ntb$ state.

The wall also exhibits a density modulation along $z$
with period $\lambda/2$. The corresponding layer spacing
is $d=1.6\,L_{\mathrm{mol}}$
(Fig.~\ref{fig:MD_variance_plots}(B)).
The density maxima exceed the mean density by approximately
$20\%$. At the midplane, they coincide with the positions
where $\hat{\bm n}\parallel\hat{\bm z}$.
In contrast, the bulk density is uniform.

Splay, bend, and twist coexist within the interfacial region.
We denote this structure by $\Ssbt$, the density-modulated
counterpart of $\Nsbt$~\cite{Drzazga2026PRE}.
This designation characterizes local interfacial order
rather than an established equilibrium smectic-like phase.

Twist is locally enhanced within the wall. At successive
density maxima, separated by $\lambda/2$, its magnitude
exceeds the bulk value and its sign alternates.
The wall thus contains a sequence of microdomains of
alternating handedness along $z$. The zero-twist surface,
$\hat{\bm n}\cdot(\nabla\times\hat{\bm n})=0$,
does not coincide with the wall midplane. Instead, 
it forms a sheet that undulates with period $\lambda$
[white contour in Fig.~\ref{fig:MD_maps}(L)], distinct from the
undulating surface on which $n_y$ vanishes.
Such undulating zero-twist surfaces are discussed further
in Sec.~S1 of the Supplemental Material.
The spatially averaged polarization vanishes on both this
sheet and the wall midplane. Saddle-splay is likewise localized
at the wall, where it varies periodically along $z$,
and vanishes in the ideal bulk $\Ntb$ state.

At $\chi=110^\circ$, both domains persisted throughout the
$5\times10^7$-step MD production run and the
$5\times10^8$-cycle MC runs
(Fig.~\ref{fig:domains_snap}).
We also varied the bend angle in increments of $5^\circ$.
At all tested state points with $\chi\geq125^\circ$,
closer to the $\Ntb$--$N$ boundary
(Fig.~\ref{fig:pd}), one domain disappeared within
$2\times10^7$ MD time steps. 
These observations characterize the persistence of locally
equilibrated two-domain configurations in finite simulation
cells. They indicate shorter lifetimes for less bent particles
under the simulated conditions.

\textit{Minimal director model---} To examine the properties of the interface, we introduce
an interpolation between the two bulk heliconical states
of opposite handedness. In this description, we measure
$x$ from the wall midplane, located at
$x/L_{\mathrm{mol}}\simeq8.5$ in the simulation coordinates. 
We write
\begin{equation}\label{eq:ansatz}
\hat{\bm n}
=
\bigl(a\cos\alpha,\ af(x)\sin\alpha,\ n_z\bigr),
\end{equation}
where $a=\sin\theta_0$. The axial component
\[
n_z
=
\left[
1-a^2\bigl(\cos^2\alpha+f^2(x)\sin^2\alpha\bigr)
\right]^{1/2}
\]
ensures that $|\hat{\bm n}|=1$.

The profile $f(x)$ describes the sign reversal of the $n_y$
modulation across the interface. We choose $f$ to be smooth,
odd, and monotonically increasing, with bulk limits
$f(\pm\infty)=\pm1$. Its slope is positive at the center,
$f'(0)>0$, and tends to zero in the bulk,
$f'(x)\to0$, as $x\to\pm\infty$.
A simple choice satisfying these conditions is
$f(x)=\tanh(x/\xi)$, where $\xi>0$ sets the characteristic
length scale of the profile. For $0<a<1$, the resulting
director field is smooth and approaches the bulk heliconical
states $\hat{\bm n}_\pm$ as $x\to\pm\infty$.

At the center of the interpolation, $x=0$, $n_y=0$, and the director 
lie in the $xz$ plane. Its tilt angle $\psi$ relative
to $\hat{\bm z}$ satisfies $\sin\psi=a\cos\alpha$.
The angle oscillates between $-\theta_0$ and $\theta_0$
with period $\lambda$. Unlike in the bulk heliconical
states, the inclination to the helix axis is therefore
not constant.

In this model, at the midplane, the splay is
\[
\left.\nabla\cdot\hat{\bm n}\right|_{x=0}
=
\frac{q_0a^2\sin(2\alpha)}{2n_z(0,z)},
\]
while the twist is
\begin{equation}\label{eq:twist}
\begin{aligned}
\tau(0,z)
&=
\left.n_z\,\partial_x n_y\right|_{x=0}\\
&=
a f'(0)n_z(0,z)\sin\alpha.
\end{aligned}
\end{equation}
The transverse derivative $\partial_x n_y$ thus generates
nonzero twist even though $n_y=0$ at the midplane.

Along the midplane, the twist changes sign every half pitch.
Its maximum magnitude, $a f'(0)$, occurs where
$\hat{\bm n}\parallel\hat{\bm z}$ and exceeds the bulk
magnitude $q_0a^2$ when $f'(0)>q_0a$.
Since $f'(0)>0$ and $n_z(0,z)>0$, the zero-twist set
intersects the midplane only at $\sin\alpha=0$,
where $|\psi|=\theta_0$.

Away from the midplane, Eq.~\eqref{eq:ansatz} gives
\begin{equation}\label{eq:twist_full}
\tau(x,z)
=
a\left[
f'(x)\sin\alpha\,\frac{1-a^2\cos^2\alpha}{n_z}
-aq_0f(x)
\right],
\end{equation}
which reduces to Eq.~\eqref{eq:twist} at $x=0$ and to the bulk
values $\tau_\pm$ as $x\to\pm\infty$. The first term is the
transverse gradient of $n_y$ across the interface; the second is
the helical winding along $\hat{\bm z}$. The zero-twist surface
is the locus where the two cancel,
\begin{equation}\label{eq:zerotwist}
aq_0f(x)
=
f'(x)\sin\alpha\,\frac{1-a^2\cos^2\alpha}{n_z}.
\end{equation}
Only the interfacial term is modulated along the helix axis, so
the cancellation point $x^\ast(z)$ shifts with $z$ and the
surface undulates with period $\lambda$. The undulation therefore follows
from Eq.~\eqref{eq:ansatz} alone, independently of any
displacement of the interpolation centre.

For any smooth, $y$-independent unit director field, the
saddle-splay invariant reduces to
$\mathcal S=(\nabla\cdot\hat{\bm n})^2-\partial_i n_j\partial_j n_i
=2(\partial_x n_x\partial_z n_z-\partial_z n_x\partial_x n_z)$.
Normalization requires $\mathcal S=0$ wherever $n_y=0$.
Differentiating $|\hat{\bm n}|=1$ along $x$ and $z$ shows
that the two projected derivatives in the $xz$ plane are
orthogonal to the same nonzero in-plane director.
They are therefore linearly dependent, and the Jacobian
vanishes. Variations of the local tilt cannot remove
this constraint.

In the simulations, the value calculated from the same reconstructed
director field on a nominal wall midplane is nonzero, and therefore
indicates a departure from these assumptions. Since the mean $n_y$ does
not vanish on the plane $x/L_{\mathrm{mol}}\simeq8.5$, that plane is
not the surface on which the director is exactly planar, and
$\mathcal S=0$ is not expected there. The constraint applies instead
to the undulating surface on which $n_y$ vanishes. That surface is
reproduced by the generalization $f(x-h(z))$ of
Eq.~\eqref{eq:ansatz}, in which the interpolation centre follows
$x=h(z)$. The zero-twist surface is a distinct sheet: on $x=h(z)$
the twist vanishes only at isolated points, so the two surfaces
undulate with the same period but intersect rather than coincide
(Sec.~S2 of the Supplemental Material).

For the interpolation in Eq.~\eqref{eq:ansatz},
$\mathcal S=-2q_0a^3ff'\sin^3\alpha/n_z$.
This expression vanishes at the interpolation centre,
is odd in $x$, and changes sign after a half-pitch
translation along $z$.

The polar field can be interpolated using the same profile,
$\bm p=p_0(\sin\alpha,\ -f(x)\cos\alpha,\ 0)$.
This expression approaches the bulk fields $\bm p_\pm$
as $x\to\pm\infty$. At the centre of the interpolation,
$x=0$, it reduces to
$\bm p=p_0\sin\alpha\,\hat{\bm x}$.
The polarization on this plane points along
$\pm\hat{\bm x}$ and has zero mean over one pitch.
Its magnitude is maximal where
$\hat{\bm n}\parallel\hat{\bm z}$.
These features are consistent with the dominant transverse
polar modulation in Fig.~\ref{fig:MD_maps}.

Within the interface, the polar field is generally not
perpendicular to the director, since
$\hat{\bm n}\cdot\bm p=p_0a(1-f^2)\sin\alpha\cos\alpha$.
However, the interpolation imposes $p_z=0$ everywhere
and therefore does not reproduce the axial polarization
observed in the simulations. Neither the density modulation
nor the equilibrium interfacial width is determined by
these interpolations.

\textit{Discussion---}
The chirality-reversing interface is a distinct density-modulated
splay--bend--twist locally stable structure, $\Ssbt$, in which an undulating surface of
vanishing twist is embedded. Splay and bend persist on this surface,
giving a local distortion pattern reminiscent of the $\Smsb$ phase that
this model forms in bulk at higher packing fractions
(Fig.~\ref{fig:pd}). The wall thus locally resembles its neighbor on the
phase diagram, but only on a thin sheet.

A twist-free director field need not be confined to a fixed
plane. Any smooth scalar field $\phi(\bm r)$ with
$\nabla\phi\neq0$ defines a twist-free director
$\hat{\bm n}=\pm\nabla\phi/|\nabla\phi|$,
normal to the level surfaces $\phi=\mathrm{const}$.
Conversely, the Frobenius theorem implies that a smooth
director field whose twist vanishes throughout an open
neighborhood can locally be written in this form.
Vanishing twist on a single surface, as observed here,
does not establish such a representation in a surrounding
neighborhood.

Let $\kappa_1$ and $\kappa_2$ denote the principal curvatures
of a level surface, with mean curvature
$H=\tfrac12(\kappa_1+\kappa_2)$ and Gaussian curvature
$K=\kappa_1\kappa_2$. We use the convention that an
outward-oriented sphere has negative principal curvatures.
The splay is then $\nabla\cdot\hat{\bm n}=-2H$ and the
saddle-splay invariant is $\mathcal S=2K$
~\cite{SelingerLCR2018}.
Such fields can support nonplanar splay--bend distortions.
Nonzero bend, however, requires the surface normal to vary
along its own direction and does not follow from surface
curvature alone (see Sec.~S1 of the Supplemental Material).

Opposite bulk handedness means that $\tau$ has opposite
signs in the two domains. For a smooth director field,
$\tau$ must therefore pass through zero across the interface.
This condition does not determine whether twist vanishes
only on a surface or throughout a layer of finite thickness.
For example, allowing the profile $f$ in
Eq.~\eqref{eq:ansatz} to remain zero over a finite interval
produces a twist-free layer of the same thickness.

The confinement of the zero-twist region to a sheet at
$\eta=0.320$ may have a thermodynamic origin.
If a finite layer had the bulk structure of $\Smsb$,
it would carry a bulk free-energy penalty on the $\Ntb$
side of the $\Ntb$--$\Smsb$ transition.
At fixed interfacial area, this penalty increases linearly
with the layer thickness. Whether such a layer 
and it's equilibrium form depends on the balance
between bulk and interfacial free energies, including
the interaction between the two boundaries of the layer.

Experimentally, the near-planar splay--bend modulation at the wall midplane
should carry the $\Ntb$ pitch and cone angle, accompanied by a density
wave of period $\lambda/2$. That such a structure forms in a purely
repulsive model shows that molecular packing alone can generate coupled
density and orientational modulations at a boundary between opposite
heliconical handedness, suggesting that this interfacial architecture
may be generic for bent-core mesogens.

\textit{Code and datasets availability---}
The source code of an original RAMPACK simulation package used to perform
Monte Carlo sampling is available at \url{https://github.com/PKua007/rampack}. 
The input script for LAMMPS and RAMPACK along with the datasets generated 
during and/or analyzed during the current
study are available from 
S.D. and P.K. upon reasonable request. \\

\acknowledgements

The authors acknowledge the support of the National 
Science Centre in Poland grant no. 2021/43/B/ST3/03135. 
Numerical simulations were carried out with the support 
of the Interdisciplinary Center for Mathematical and Compu-
tational Modeling (ICM) at the University of Warsaw
under grant no. G27-8.

\bibliography{main}

\ifincludeSM
\clearpage
\begin{center}
\textbf{\large Supplemental Material}
\end{center}

\setcounter{equation}{0}
\setcounter{figure}{0}
\renewcommand{\theequation}{S\arabic{equation}}
\renewcommand{\thefigure}{S\arabic{figure}}
\renewcommand{\theHequation}{S\arabic{equation}}
\renewcommand{\theHfigure}{S\arabic{figure}}

\section*{S1. Twist-free director fields and leaf geometry}

Splay--bend structures of zero twist can be much more complex than planar ones. The
zero-twist condition implies an orthogonal curved-surface foliation: any sufficiently
smooth scalar field $\phi(\bm r)$ with $\nabla\phi(\bm r)\neq0$ defines a possible
splay--bend director field
\begin{equation}\label{eq:S-gradient}
\hat{\bm n}(\bm r)=\pm\frac{\nabla\phi(\bm r)}{\|\nabla\phi(\bm r)\|}.
\end{equation}
The converse also holds. Let $\Omega\subset\mathbb{R}^{3}$ be compact and let the director
field be $C^{1}$ and twist-free, $\tau=\hat{\bm n}\cdot(\nabla\times\hat{\bm n})=0$. Then:
\begin{itemize}
\item \emph{Local leaves.} For every $p\in\Omega$ there exist a neighbourhood $U\ni p$ and a
$C^{2}$ function $\phi:U\to\mathbb{R}$ with $\nabla\phi\neq0$ on $U$ such that
Eq.~\eqref{eq:S-gradient} holds on $U$; the level sets $\{\phi=\mathrm{const}\}$ are pairwise
disjoint $C^{2}$ surfaces in $U$, everywhere orthogonal to $\hat{\bm n}$.
\item \emph{Orthogonal curved-surface foliation on $\Omega$.} These local level sets patch
together to a codimension-one $C^{2}$ foliation of $\Omega$ (leaves may end at
$\partial\Omega$). Through every point of $\Omega$ passes a unique maximal connected surface
everywhere orthogonal to $\hat{\bm n}$.
\item \emph{Regularity.} If $\hat{\bm n}\in C^{r}$ with $r\ge1$, the leaves are of class
$C^{r+1}$.
\item \emph{Curvature identities on the leaves.} On any leaf with unit normal $\hat{\bm n}$,
let $\kappa_{1},\kappa_{2}$ be the principal curvatures, $H=\tfrac12(\kappa_{1}+\kappa_{2})$
the mean curvature, and $K=\kappa_{1}\kappa_{2}$ the Gaussian curvature. At every regular point,
\begin{equation}\label{eq:S-curvatures}
\nabla\cdot\hat{\bm n}=-2H,\qquad \tau=0,\qquad \mathcal S=2K,
\end{equation}
and the bend vector $\bm\beta=(\hat{\bm n}\cdot\nabla)\hat{\bm n}$ has magnitude equal to the
Frenet curvature of the director lines. The curvatures are those of the shape operator
$-\mathrm{d}\hat{\bm n}$; e.g.\ for a sphere of radius $r$ with outward normal,
$\kappa_{1}=\kappa_{2}=-1/r$ and $\nabla\cdot\hat{\bm n}=2/r$.
\end{itemize}

For any unit field $\hat{\bm n}$ the saddle-splay density is
\begin{equation}\label{eq:S-saddle}
\begin{split}
\mathcal S&=\nabla\cdot\big[\hat{\bm n}\,(\nabla\cdot\hat{\bm n})+\hat{\bm n}\times(\nabla\times\hat{\bm n})\big]\\
&=(\nabla\cdot\hat{\bm n})^{2}-\partial_{i}n_{j}\,\partial_{j}n_{i}.
\end{split}
\end{equation}
Pointwise, $\mathcal S=2\det\mathsf S+\tau^{2}/2$, where $\mathsf S$ is the symmetric part of
the $2\times2$ matrix $\partial_a n_b$ with $a,b$ spanning the plane normal to $\hat{\bm n}$;
for a twist-free field $\det\mathsf S=K$ and Eq.~\eqref{eq:S-curvatures} is recovered.
For fields independent of $y$, as the averaged maps of the main text, all terms of
Eq.~\eqref{eq:S-saddle} containing $\partial_y$ vanish, and with
$\nabla\cdot\hat{\bm n}=\partial_x n_x+\partial_z n_z$ one obtains
\begin{equation}\label{eq:S-saddle2d}
\mathcal S=2\,(\partial_x n_x\,\partial_z n_z-\partial_z n_x\,\partial_x n_z).
\end{equation}

The theorem requires $\tau=0$ on an open set, as in bulk $\Smsb$. On the zero-twist surface of
the wall, $\tau$ vanishes only on the surface itself, so the leaf construction does not apply
there; the pointwise identity still gives $\mathcal S=2\det\mathsf S$ on it.

\section*{S2. Periodic chiral domain wall}

Let the wall centre be periodic in the $(x,z)$ plane, $x=h(z)$ with $h(z+\lambda)=h(z)$
(e.g.\ $h(z)=h_{0}\cos(q_{0}z+\delta)$), define the signed distance
\begin{equation}
s(x,z)=x-h(z),
\end{equation}
and choose a smooth odd profile $f(s)$ with $f(\pm\infty)=\pm1$ and $f(0)=0$. Because the two
monodomains differ only in the sign of the $y$ components of $\hat{\bm n}$ and $\bm p$, the
interpolation acts on these components only:
\begin{equation}\label{eq:S-wall}
\begin{split}
\hat{\bm n}&=\big(a\cos\alpha,\ af(s)\sin\alpha,\ n_z\big),\\
\bm p&=p_0\big(\sin\alpha,\ -f(s)\cos\alpha,\ 0\big),
\end{split}
\end{equation}
with $a=\sin\theta_0$, $\alpha=q_0z+\phi_0$ and $n_z$ fixed by $|\hat{\bm n}|=1$, so that
$\hat{\bm n}\to\hat{\bm n}_\pm$ and $\bm p\to\bm p_\pm$ as $s\to\pm\infty$. (Multiplying the
whole of $\bm p$ by $f(s)$ would instead give $-\bm p_+$, not $\bm p_-$, as $s\to-\infty$,
since it also reverses $p_x$. The interpolation preserves $\bm p\perp\hat{\bm n}$ only in the
bulk; this does not affect the director results.)

On the wall centre $s=0$, $\hat{\bm n}=(a\cos\alpha,\,0,\,n_z)$ has planar
splay--bend order, and $\bm p=p_0\sin\alpha\,\hat{\bm x}$ lies in the same plane with zero
pitch average: the sheet is globally, though not locally, non-polar. The twist on it is
\begin{equation}\label{eq:S-twist}
\tau\big|_{s=0}=a\,f'(0)\,\sin\alpha\,\big[n_z+h'(z)\,n_x\big].
\end{equation}
Hence for $f'(0)=0$ the sheet $s=0$ is twist-free: an achiral sheet at the wall centre. For
$f'(0)\neq0$, as in the simulations, twist concentrates on the wall and vanishes on $s=0$ only
where $\sin\alpha=0$ or $n_z+h'n_x=0$; the zero-twist surface then departs from $s=0$,
crossing it at these points. For $h\equiv0$, Eq.~\eqref{eq:S-twist} reduces to
Eq.~\eqref{eq:twist} of the main text.

\section*{S3. Bend-induced form of the polarization}

The transverse polarization of the heliconical monodomains may equivalently be written as
\begin{equation}
\begin{split}
\bm p_{\pm}&=\zeta\,\hat{\bm n}_{\pm}\times(\nabla\times\hat{\bm n}_{\pm})
=-\zeta\,(\hat{\bm n}_\pm\cdot\nabla)\hat{\bm n}_\pm\\
&=\zeta\,q_{0}\sin\theta_{0}\cos\theta_{0}\,\big(\sin\alpha,\ \mp\cos\alpha,\ 0\big),
\end{split}
\end{equation}
with a material coefficient $\zeta>0$, so that $p_{0}=\zeta\,q_{0}\sin\theta_{0}\cos\theta_{0}$.

\section*{S4. MD starting configuration}

\begin{figure}[htb]
    \centering
    \includegraphics[width=0.8\linewidth]{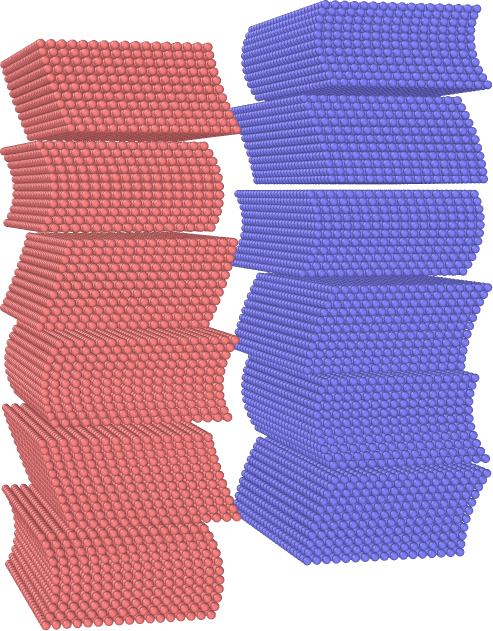}
    \caption{Starting configuration for $\eta=0.320$, $\chi = 110^\circ$ (MD). For clarity,
    only about 1/8 of the molecules is shown. One lattice monodomain was created first; the
    second domain was obtained by mirroring it in the interface plane and shifting it along
    $z$ by half of the equilibrated $\Ntb$ pitch. Colouring according to
    $(\hat{\bm n}\times\bm p)_z$, with $\hat{\bm n}$ oriented such that $n_z>0$.}
    \label{fig:MD_starting_configuration}
\end{figure}

\fi

\end{document}